\documentclass[aps,prd,twocolumn,superscriptaddress,nofootinbib,preprintnumbers]{revtex4-2}
\usepackage{amsmath}
\usepackage{graphicx}
\usepackage{subfigure}
\usepackage{amssymb}
\usepackage{xcolor}
\usepackage{multirow}
\usepackage{cancel}
\usepackage{color}
\usepackage{ulem}
\usepackage{listings}
\usepackage{xcolor}
\usepackage{float}
\usepackage{slashed}
\usepackage{wrapfig}
\usepackage{mathtools}
\usepackage[colorlinks,citecolor=blue]{hyperref}
\usepackage[T1]{fontenc}  
\usepackage[utf8]{inputenc}  
\usepackage{times}
\begin{document}

\title{\boldmath The mass-coupling  effect in  leptogenesis}

\author{Shinya Kanemura}
\email{kanemu@het.phys.sci.osaka-u.ac.jp}
\affiliation{Department of Physics, The University of Osaka, Toyonaka, Osaka 560-0043, Japan}

\author{Shao-Ping Li}
\email{lisp@het.phys.sci.osaka-u.ac.jp}
\affiliation{Department of Physics, The University of Osaka, Toyonaka, Osaka 560-0043, Japan}
 
\begin{abstract}
	Particle  decay in leptogenesis  provides a simple avenue to explain the baryon asymmetric universe, where  the decaying particle  can provide the out-of-equilibrium condition to create a net lepton  asymmetry. It is widely anticipated that the lepton asymmetry would  be changed significantly  by varying   couplings and the decaying particle mass, especially  in the weak washout regime.  Contrary to this naive expectation,  we demonstrate  a general phenomenon in a class of leptogenesis scenarios from heavy particle decay, where varying the  mass and  couplings would not modify the lepton asymmetry in a noticeable way, as these  mass and coupling effects are largely canceled out from the  evolution of the decaying particle.  It points out that a much broader parameter space in the  mass and couplings will open automatically once leptogenesis is realized in a benchmark point; however, tuning the mass and couplings to boost leptogenesis   will be challenging.
\end{abstract}

\maketitle


\preprint{OU-HET 1278} 
	\section{Introduction}
	\label{sec:intro}
The matter-antimatter asymmetry observed in our current  universe~\cite{Planck:2018vyg}, or the  baryon asymmetry of the universe (BAU), has  raised an important  question  in the standard model (SM)  for the dynamical origin.  Over the past decades, leptogenesis through heavy particle decay~\cite{Fukugita:1986hr,Luty:1992un,Covi:1996wh} has become one of the leading  mechanisms to explain the BAU problem,  where the necessary out-of-equilibrium  condition is generally provided by the decaying particle itself. When the cosmic temperature drops   below the mass of the decaying particle, production processes of the decaying particle can no longer compete with depletion processes, leading to the  decaying  particle falling out of thermal equilibrium.

Once the nonthermal condition is realized in leptogenesis,   a net CP asymmetry  and consequently  a net lepton-number asymmetry  can be induced and reprocessed  into the baryon asymmetry via electroweak sphaleron processes at finite temperatures~\cite{Kuzmin:1985mm}.  In most leptogenesis,  the lepton  asymmetry depends on the imaginary part of Yukawa couplings, the nonthermal distribution function of the decaying particle, and the inverse mass difference if the CP asymmetry is induced by quantum self-energy corrections.  These overarching features have established common platforms to effectively enhance the lepton  asymmetry.

\textit{The mass  effect.} If the  decaying particle  itself provides a nonthermal condition in leptogenesis, then its mass cannot be arbitrarily small. In principle, it should be  larger than the sphaleron decoupling temperature, $T_{\rm sph}\approx 132$~GeV~\cite{DOnofrio:2014rug}.   If the decaying particle with an electroweak-scale mass was once in  thermal contact with the plasma,  its  distribution function would then have  a small  departure from thermal equilibrium,  thereby suppressing the generation of the lepton  asymmetry. For leptogenesis from particle decay, one can anticipate  that the    lepton  asymmetry  would depend on the decay rate and hence on the decaying particle mass, though in some nontrivial ways. This can be seen from the Boltzmann equation of the lepton asymmetry, where the CP-violating source term is proportional to the decay rate; see e.g., Refs.~\cite{Buchmuller:2004nz,Davidson:2008bu} for review. Therefore, varying  the decaying particle mass may be able to  enhance the final lepton asymmetry.  

\begin{figure}[t]
	\centering
	\includegraphics[width=0.25\textwidth]{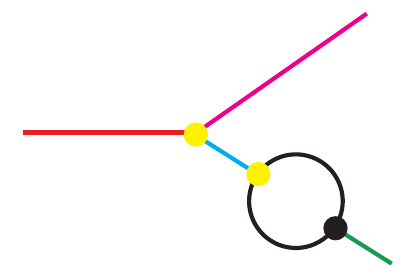}
	\caption{\label{fig:cp-eh} The typical  topology showing Yukawa coupling enhancement of leptogenesis from particle decay. The black blob denotes the small coupling that corresponds to the out-of-equilibrium decay product (green), and the yellow blobs denote large couplings associated with the thermalized flavor (cyan) which  enhances the lepton asymmetry.}
\end{figure}

\textit{Large Yukawa couplings.} It should be mentioned that the nonthermal condition  is not necessarily  given  by the decaying particle. 
Instead,  it can be provided by decay products. In this case, the decay products may  be absent  initially in the plasma and never reach thermal equilibrium.  Generally,   this   corresponds to the so-called weak washout regime~\cite{Buchmuller:2005eh}, with the typical examples given by freeze-in leptogenesis~\cite{Akhmedov:1998qx,Dick:1999je,Asaka:2005pn,Li:2020ner,Goudelis:2021qla}. Nevertheless,  the Yukawa couplings in this case must be restricted to small values in order to maintain the out-of-equilibrium condition.   Also in this case, increasing the Yukawa couplings  to larger values is challenging since both the CP-violating source and the washout rate will be enhanced. The net effect can hardly be  anticipated analytically, as  solutions to  the notorious  integro-differential Boltzmann equations  require dedicated numerical analysis. 

Nevertheless, if the nonthermal condition is partly  given by the decaying   particle, it becomes possible to make some flavors of the decay products in thermal equilibrium and the other flavors out of equilibrium. In this case,  the Yukawa couplings associated with the thermalized flavors can  be larger in general. As a result, it features a hierarchy of couplings in the prediction of the final lepton asymmetry,  potentially realizing a parameter enhancement mechanism~\cite{Hambye:2001eu,Li:2021tlv}. The typical topology that features such a  parameter enhancement is shown in Fig.~\ref{fig:cp-eh}.

\textit{Mass degeneracy.} If the CP asymmetry is induced by one-loop self-energy corrections, the final lepton asymmetry would be proportional to the inverse of the  mass difference from the two particles attached to the self-energy loop. In this case, a small mass difference due to these quasi-degenerate particles will lead to resonant enhancement of the CP asymmetry, as  discussed  in  vacuum resonant leptogenesis~\cite{Pilaftsis:2003gt,Pilaftsis:2005rv} and vacuum-forbidden resonant leptogenesis~\cite{Garbrecht:2012pq,Kanemura:2024dqv,Kanemura:2024fbw}.

The aforementioned mass  effect, large Yukawa couplings, and mass degeneracy are nowadays the most common  practice in opening up  parameter space to boost   leptogenesis.  Especially, they can lower down the scale of leptogenesis, such that new particles introduced beyond the SM can be as low as in the TeV or electroweak scale, making   direct production attainable at colliders.  However, is it always the case that these enhancement mechanisms work efficiently to boost leptogenesis? For mass degeneracy, it has been noticed for a while that resonant enhancement from vacuum mass degeneracy would  be suppressed  by thermal mass and width effects~\cite{Pilaftsis:1997jf,Garny:2011hg,Hohenegger:2014cpa}, thereby limiting  the maximal effect caused by the resonant enhancement. Yet, there is still no clear and simple analysis of the enhancement from the mass-coupling effects.

In this work, we elaborate on   the mass-coupling effects on a generic class of leptogenesis, where the decaying particle and some decay products provide the nonthermal condition.  Contrary to naive expectations, we find that the lepton  asymmetry is basically  independent of the  Yukawa couplings associated with thermalized decay products,  and of the decaying particle mass.  Therefore,  varying  the Yukawa couplings and the  decaying particle mass will  not increase the final lepton  asymmetry in a noticeable way.  The  reason for such   independence lies in  the large cancellation effect from  the evolution of the decaying particle, where increasing (decreasing) the Yukawa couplings (the mass) will lead to a suppressed  nonthermal  distribution function of the decaying particle. While the evolution of the  distribution function from the decaying particle depends nontrivially on    the mass and Yukawa couplings, the fact that   culmination of leptogenesis at the cosmic  temperature   below the decaying particle mass  will  turn  the  nontrivial dependence into  a  simple scaling regime.

\section{General mass-coupling dependence}
\label{sec:app}
For generality, let us first discuss in this section some common features of the mass and couplings in the generation of the lepton asymmetry. Throughout the discussions, we will focus on  the typical leptogenesis scenarios with a CP-violating source induced at quartic order of Yukawa couplings, and with a weak washout regime where the washout effect can be neglected as a good approximation. In this circumstance,   the Boltzmann   kinetic equation   reduces to a simple integral equation:
\begin{align}\label{eq:Yell}
	Y_{\ell}&\equiv\frac{n_\ell-\bar n_\ell}{s}
=\int \frac{\mathcal{S}_{\rm CP}}{s H T} dT
		\nonumber \\[0.2cm]
		&= c_{\rm md}\text{Im}[y^{\prime 2} y^2]\left(\frac{M_{\rm Pl}}{m}\right)\int z^4 \mathcal{\bar S}_{\rm CP}(z) dz\,,
\end{align}
where $n_{\ell}, \bar n_\ell$ are the number densities of leptons and antileptons,  $s$ is the entropy density in the SM plasma, and $H$ is the Hubble parameter. The dimensionless  constant $c_{\rm md}$ in Eq.~\eqref{eq:Yell} is introduced to absorb  model-dependent parameters, such as phase-space factors and  constants from squared amplitudes,  which   are independent of the  mass and  Yukawa couplings.  The Planck mass  $M_{\rm Pl}$ appears due to the  temperature  integration under the cosmic expansion.  It should be mentioned that the  general parameterization given in Eq.~\eqref{eq:Yell} can be readily  spelled out to match the model-specific formulas; see e.g., Refs.~\cite{Giudice:2003jh,Buchmuller:2004nz,Davidson:2008bu}. We should emphasize, however,   that Eq.~\eqref{eq:Yell} cannot be applied to scenarios where the washout processes are important, such as the canonical leptogenesis based on type-I seesaw~\cite{Fukugita:1986hr,Luty:1992un,Covi:1996wh}.

 In the second line  of Eq.~\eqref{eq:Yell}, we   extracted the overall mass scale from the CP-violating source
 \begin{align}\label{eq:SCP-norm}
\mathcal{S}_{\rm CP}\equiv m^4 \mathcal{\bar S}_{\rm CP}\,,
 \end{align}
 by using the dimensionless time variable $z\equiv m/T$, where $m$ denotes the decaying particle mass. Such factorization is generally expected in leptogenesis when $m\gtrsim T$, as the energy scale is then dominated by the decaying particle mass.  
 
 The imaginary part of quartic Yukawa couplings is shown explicitly, where $y'$ denotes the  Yukawa vertex associated with the nonthermal decay product (denoted as $\chi'$)  while $y$ denotes the Yukawa vertex associated with  the  self-energy loop or equivalently the one associated with the thermalized decay products (denoted as $\chi$).  Note that since $y$ should be large enough to keep the $\chi$ flavors in thermal equilibrium,  it can  also affect the  evolution of the decaying particle.  In general, one can adjust both $y'$ and $y$ to enhance $Y_\ell$. However, an upper bound of  $y'$ exists since $y'$ is associated with the  nonthermal $\chi'$ flavor.   The room for varying  $y$ is generally much larger, spanning orders of magnitude provided that  this Yukawa vertex keeps the associated $\chi$ flavors in thermal equilibrium. Therefore, we would  expect that increasing $y$ is more effective to enhance $Y_\ell$. In section~\ref{sec:evolution}, however, we will show that this is not the case, as the integration  over $\mathcal{\bar S}_{\rm CP}$ would be suppressed by a  larger $y$, basically  canceling  out the  increase of $Y_\ell$ from $\text{Im}(y^{\prime 2}y^2)$.

%
Under the parameterization of Eq.~\eqref{eq:Yell}, certain model dependence still arises from the CP-violating source $\mathcal{\bar S}_{\rm CP}$, which is  a multi-dimensional momentum integration of distribution functions. Nevertheless, for the decaying particle providing  the nonthermal condition, there is a generic scaling 
\begin{align}\label{eq:SCP-deltaf}
\mathcal{\bar S}_{\rm CP}\propto \delta f\,, 
\end{align} 
where $\delta f\equiv f -f^{\rm eq}$ denotes the departure from thermal equilibrium. This is a robust feature dictated by unitarity and the CPT theorem, which states that no CP asymmetry would be generated in thermal equilibrium.  Furthermore, by extracting the overall mass scale via Eq.~\eqref{eq:SCP-norm}, the dominant dependence of $\mathcal{\bar S}_{\rm CP}$ on $m$ would originate from $\delta f$ while the mass dependence through $z$ is largely removed since leptogenesis generally culminates at a known moment $z=\mathcal{O}(1)$ when the Boltzmann suppression becomes important. Then, we expect the integration of $\mathcal{\bar S}_{\rm CP}$ is dominantly determined at this known  moment.
Note that, however, a larger mass with $z=\mathcal{O}(1)$ indicates that the end of leptogenesis is earlier, where the  time  duration of  subsequent evolution of the lepton asymmetry before sphaleron decoupling would be different. Nevertheless, the impact  of the  late-time evolution can be neglected if   the weak washout regime continues as a good approximation. 

In parameterizing the lepton asymmetry via Eq.~\eqref{eq:Yell}, we are able to analyze the mass-coupling effects of leptogenesis largely independent of specific particle physics models. In particular, the most important effects of varying the mass and Yukawa couplings would  only appear through  the common factors $m, \text{Im}[y^{\prime 2} y^2]$, and $\delta f$ inside the integration.   Next, we  calculate the evolution of $\delta f$ under different masses and Yukawa couplings $y$, and see how  the net effect   from these common factors changes.

\section{Leptogenesis from  particle decay}\label{sec:evolution}
Leptogenesis can be realized by either scalar or fermion decay. For definiteness, we will consider leptogenesis from  scalar decay~\cite{Ma:1998dx,Dick:1999je,Hambye:2000ui,Hambye:2003ka,Garbrecht:2012qv,Li:2021tlv}, but the general conclusions drawn in the following are also applicable to the fermion case.  

The relevant  Yukawa interaction  is given as 
\begin{align}\label{eq:lag}
	\mathcal{L}=- y\bar\ell \tilde{\phi}  \chi_R\,, 
\end{align}
where  $\ell$ denotes the $SU(2)_L$ lepton doublets coupling to  right-handed  gauge fermion singlets $\chi_R= (1+\gamma_5) \chi/2$, and $\tilde{\phi}=i \sigma_2 \phi^*$ with $\sigma_2$ the second Pauli matrix and $\phi$  a gauge $SU(2)_L$ doublet. 
This Yukawa interaction   will be the prototype  for leptogenesis analyzed  throughout the  work, where the CP asymmetry can be generated  by one-loop self-energy corrections either to the $\chi$ flavors~\cite{Hambye:2016sby,Hambye:2017elz} or to the $\ell$ flavors~\cite{Garbrecht:2010sz,Garbrecht:2012pq,Li:2020ner,Kanemura:2024dqv,Kanemura:2024fbw}, and can also be induced by one-loop vertex corrections~\cite{Hambye:2003ka}.  We assume the scalar has a vacuum mass $m_\phi$ and is much heavier than  the decay products. 

The evolution of the scalar distribution function $f_\phi$ can be computed by the Boltzmann equation,
\begin{align}\label{eq:Boltzmann}
	\frac{\partial f_\phi}{\partial t}-H p_\phi\frac{\partial f_\phi}{\partial p_\phi}&=-\mathcal{C}_{\rm Yuk}\,,
\end{align}
where $p_\phi\equiv |\vec p_\phi|$ and the Hubble parameter in the radiation-dominated epoch reads
\begin{align}
	H\approx 1.66\sqrt{g_\rho(T)}\frac{T^2}{M_{\rm Pl}}\,,
\end{align}
with $g_{\rho}(T)$ being  the relativistic degrees of freedom in energy density and $M_{\rm Pl}\approx 1.22\times 10^{19}$~GeV being  the Planck mass.   $\mathcal{C}_{\rm Yuk}$ denotes the the Boltzmann collision rate from decay and inverse decay $\phi\leftrightharpoons \ell+\bar\chi$.

Note that  for gauge scalars or any gauged particles, gauge  interactions should in general be included in the Boltzmann collision rates. We will neglect the gauge-scalar interactions in determination of  $f_\phi$, such that the following conclusions can be directly applied to any gauge singlet decay. It is worthwhile to mention that if  Yukawa couplings are too small such that gauge-scalar interactions  govern the evolution of $f_\phi$, then the effect of varying the small Yukawa couplings will only appear through $\text{Im}[y^{\prime 2} y^2]$.  Determining the regimes in which the Yukawa interaction supersedes the gauge interaction in governing the evolution of $f_\phi$ necessitates a more dedicated numerical analysis of the integro-differential Boltzmann equation, which we will not pursue here.

The Boltzmann collision rate $\mathcal{C}_{\rm Yuk}$ is     determined by  
\begin{align}
	\mathcal{C}_{\rm Yuk}&=\frac{1}{2E_\phi}\int d\Pi_\ell  d\Pi_{\bar \chi}(2\pi)^4\delta^4(p)|\mathcal{M}_{\phi \to\ell\bar \chi}|^2\mathcal{F}_{\rm Yuk}\,,
\end{align}
where the phase-space factor is defined as
\begin{align}
 d\Pi_i	\equiv \frac{d^3p_i}{(2\pi)^3 2E_i}\,,
\end{align} 
for $i=\ell, \bar\chi$. The statistics function reads
\begin{align}
	\mathcal{F}_{\rm Yuk}&=f_\phi(1-f^{\rm eq}_\ell)(1-f^{\rm eq}_{\bar\chi})-f^{\rm eq}_\ell f^{\rm eq}_{\bar\chi}(1+f_\phi)
\nonumber\\[0.2cm]
&	=\delta f_\phi\left[1-f^{\rm eq}_\ell-f^{\rm eq}_{\bar\chi}\right],
\end{align}
with  $\delta f_\phi\equiv f_\phi-f_\phi^{\rm eq}$ and $f_\phi^{\rm eq}(E)=(e^{E/T}-1)^{-1}$.  In addition, 
we  take thermal distribution functions for lepton doublets and the thermalized fermion singlet, 
\begin{align}
	f^{\rm eq}_{\ell}(E)=f^{\rm eq}_{\chi}(E)=\frac{1}{e^{E/T}+1}\,,
\end{align}
where  small chemical potentials are neglected.

Since  the scalar  already provides a nonthermal condition for leptogenesis, we can make  one of the fermion singlet  flavors, $\chi'$,  out of equilibrium by assuming  small Yukawa matrix elements $|y_{ij'}|$ with $i$ being the lepton flavors  and $j'$ being  the nonthermal $\chi'$ flavor. Then we  make the other $\chi$ flavors  in    thermal equilibrium with the SM plasma, where we expect  larger Yukawa matrix elements $|y_{ij}|$ with $j$ being the thermalized $\chi$ flavors.  This is where a coupling hierarchy $|y_{ij'}|\ll |y_{ij}|$ can    enhance the lepton asymmetry~\cite{Hambye:2001eu,Li:2021tlv}. 

We   approximate the perturbation of the scalar distribution function  by  a small time-dependent chemical potential
\begin{align}\label{eq:fphi-per}
	f_\phi=\frac{1}{e^{(E-\mu)/T}-1}\,, \quad 	\delta f_{\phi}= \tilde{\mu}\frac{e^{E/T}}{(e^{E/T}-1)^2}\,,
\end{align}
where $\delta f_\phi$ is derived by keeping only the leading-order perturbation, with $\tilde{\mu}\equiv \mu/T$.  

For numerical analysis, the left-hand side of Eq.~\eqref{eq:Boltzmann} can  be   simplified by defining 
\begin{align}
	z\equiv \frac{m_\phi}{T}\,, \quad r_\phi\equiv \frac{p_\phi}{T}\,.
\end{align}
With the relation $dT/dt\approx -HT$ during the radiation-dominated epoch, the left-hand side becomes a total derivative  according to 
\begin{align}
	\frac{\partial f_\phi}{\partial t}-Hp_\phi \frac{\partial f_\phi}{\partial p_\phi}=H z\frac{d f_\phi}{dz}=H z\frac{d \delta f_\phi}{dz}+H z\frac{d  f^{\rm eq}_\phi}{dz}\,,
\end{align}
where the Hubble parameter   can be rewritten as 
\begin{align}
	H(z)=  \frac{m_\phi^2}{\bar M_{\rm Pl}z^2}\,,
\end{align}
with $g_\rho\approx 110.74$ used and $\bar M_{\rm Pl}\approx 6.97\times 10^{17}$~GeV. 
 Explicitly, we have
\begin{align}
	\frac{d \delta f_\phi}{dz}&=\frac{1}{2\left[\cosh\left(\xi\right)-1\right]}\frac{d\tilde \mu}{dz}-\frac{z \coth\left(\xi/2\right)}{2\xi \left[\cosh\left(\xi\right)-1\right]}\tilde{\mu}\,,
	\\[0.2cm]
	\frac{d f^{\rm eq}_\phi}{dz}&=-\frac{z}{2\xi \left[\cosh\left(\xi\right)-1\right]},
\end{align}
where   we have defined 
\begin{align}
	\xi\equiv\sqrt{z^2+r_\phi^2}\,.
\end{align}

The tree-level squared amplitude reads
\begin{align}
	|\mathcal{M}_{\phi \to\ell\bar\chi}|^2&=|y|^2 m_\phi^2\,,
\end{align}
where 
\begin{align}\label{eq:ysq}
|y|^2\equiv \sum_{ij}|y_{ij}|^2\,,
\end{align} 
corresponds to   summation  over thermalized $\chi$ flavors and over lepton flavors.  

Applying the above ingredients, we can simplify $\mathcal{C}_{\rm Yuk}$ as
\begin{align}
	\mathcal{C}_{\rm Yuk}
	=  \frac{|y|^2 m_\phi}{8\pi}\tilde\mu(z)\mathcal{I}_{\rm Yuk}\,,
\end{align}
where  
\begin{align}
	\mathcal{I}_{\rm Yuk}\equiv \frac{z e^{\xi}}{r_\phi\xi\left(e^{\xi}-1\right)^2}\ln\left(\frac{\cosh\left[\left(\xi+r_\phi\right)/4\right]}{\cosh\left[\left(\xi-r_\phi\right)/4\right]}\right).
\end{align}
The   Boltzmann equation reduces to 
\begin{align}\label{eq:dmudz-fin}
	\frac{d\tilde{\mu}}{dz}=\frac{z}{\xi}+\mathcal{D}\times \tilde{\mu}\,,
\end{align}
where the  initial condition is given by $\tilde{\mu}=0$, and $\mathcal{D}$ denotes the damping or friction force,
\begin{align}
\mathcal{D}&=\frac{z\coth\left(\xi/2\right)}{\xi}
	-\frac{|y|^2\bar M_{\rm Pl}z}{4\pi m_\phi}\left[\cosh\left(\xi\right)-1\right] \mathcal{I}_{\rm Yuk}\,.
\end{align}
The first term in $\mathcal{D}$ is smaller than the second term  for the typical  parameter space of $m_\phi$ and  $|y|$, such that 
\begin{align}\label{eq:Dprop}
	\mathcal{D}\propto |y|^2 \left(\frac{\bar M_{\rm Pl}}{m}\right)\,.
\end{align}
 It is worth mentioning that the structure shown in Eq.~\eqref{eq:dmudz-fin} and also in Eq.~\eqref{eq:Dprop}  is  applicable to fermion singlet decay. This common  dependence dictates how the departure of the distribution function  from thermal equilibrium  will change under the variation of couplings and the decaying particle mass. 
\begin{figure*}[t]
	\centering
	\includegraphics[scale=0.301]{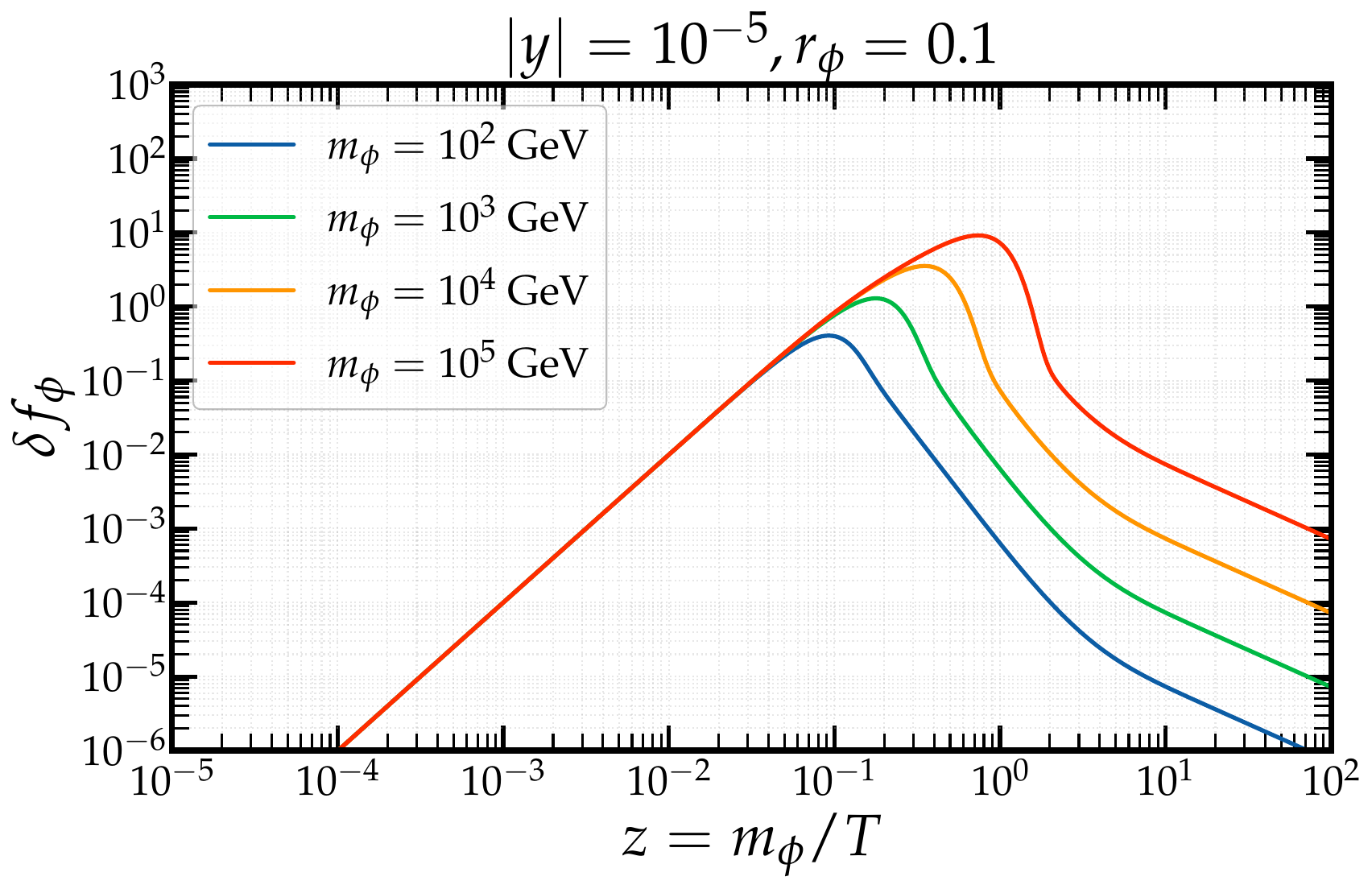} \quad
	\includegraphics[scale=0.301]{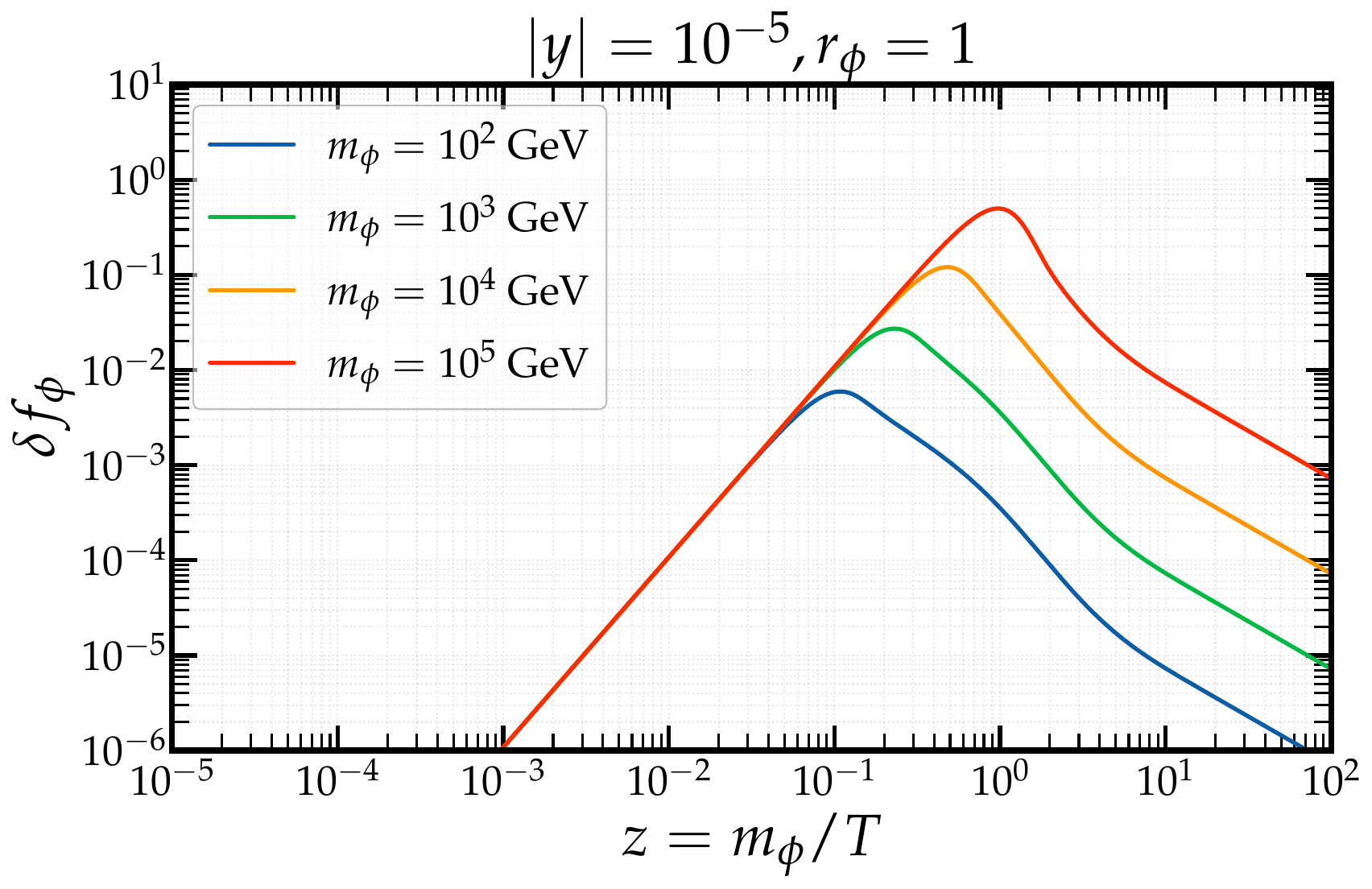}
	\\[0.5cm]
	\includegraphics[scale=0.301]{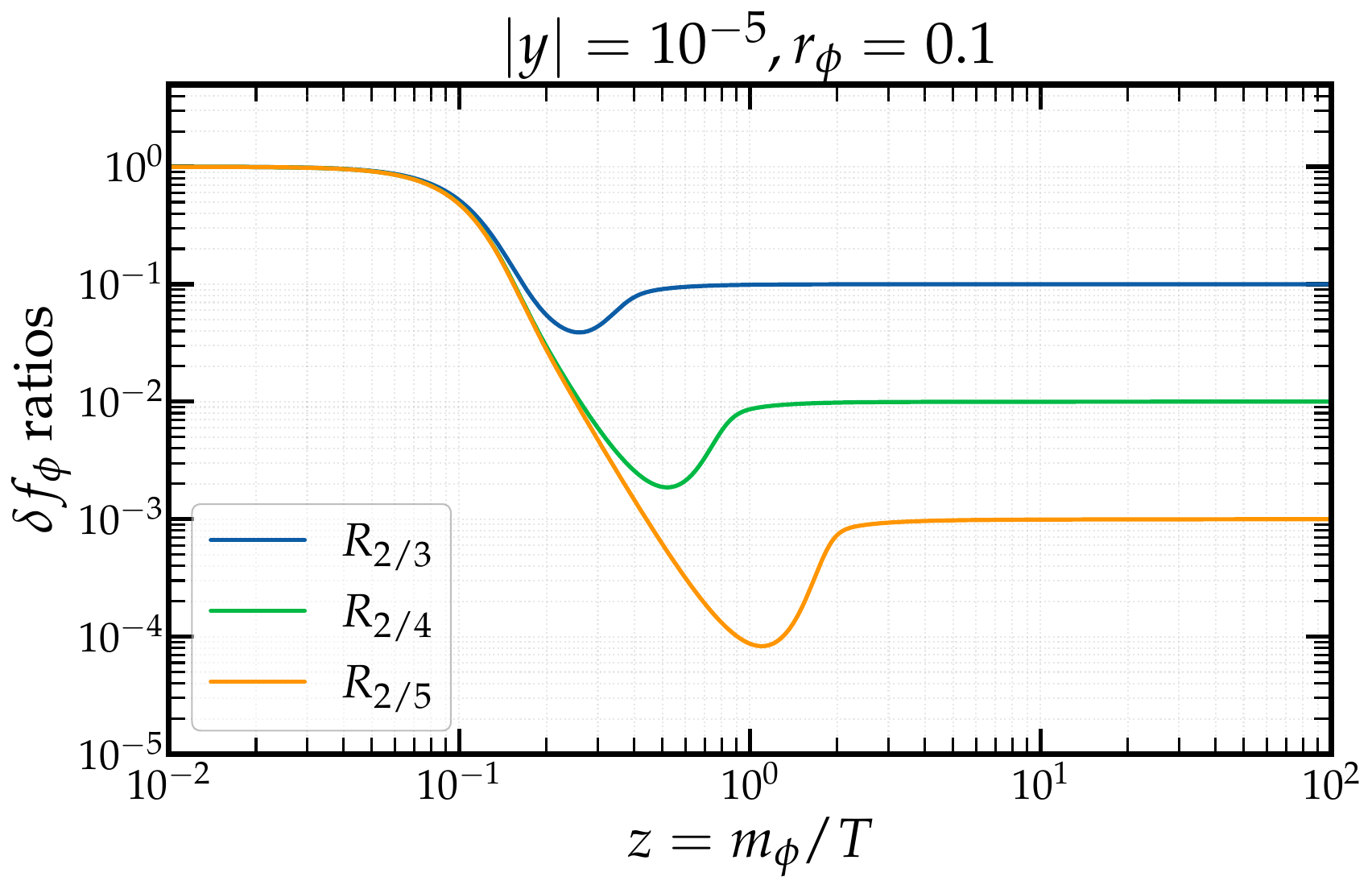} \quad
	\includegraphics[scale=0.301]{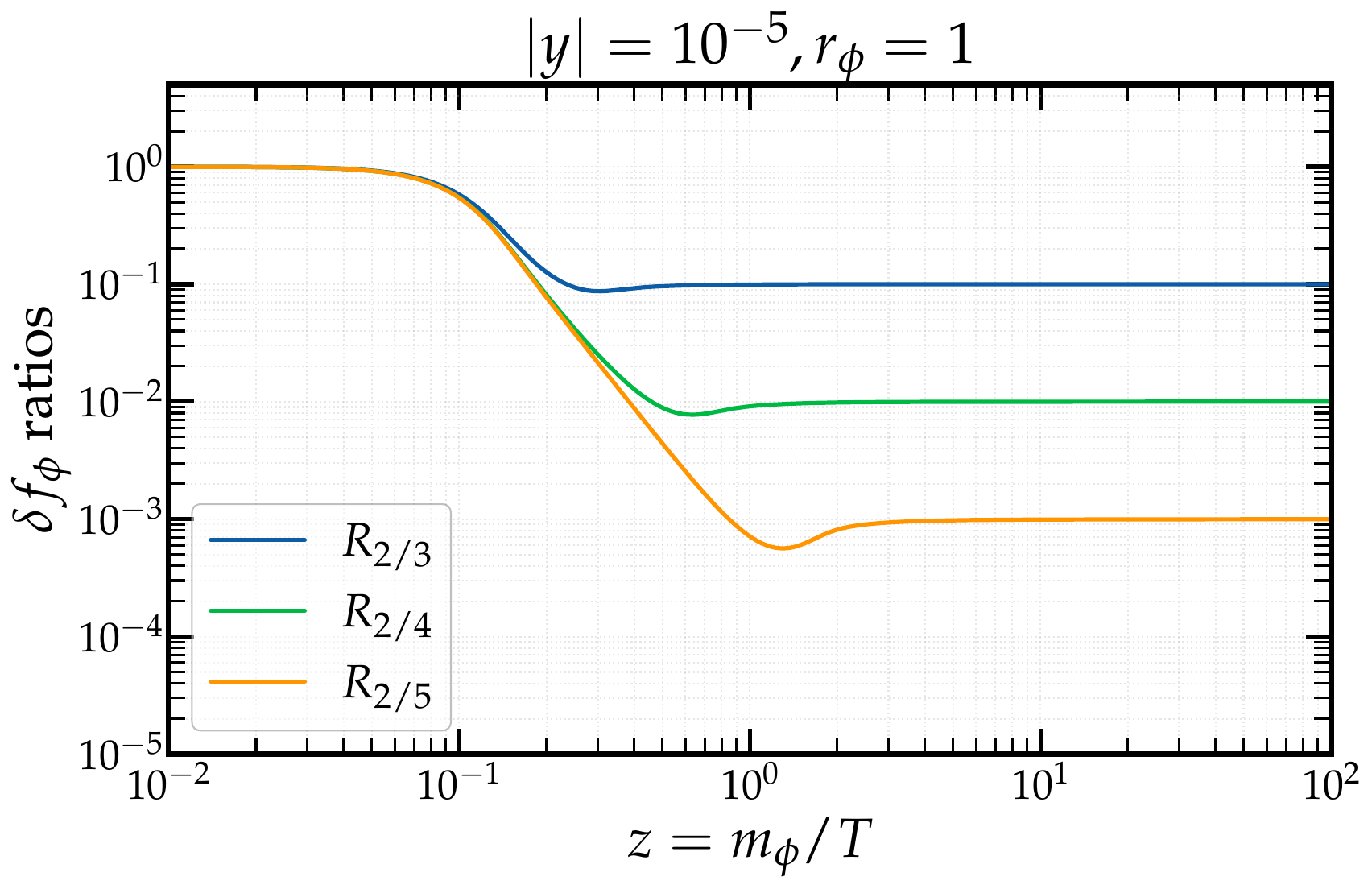}
	\caption{\label{fig:fphi-m}Top: the evolution of $\delta f_\phi$ under the variation of the decaying particle mass.  The Yukawa couplings are fixed at $10^{-5}$ for reference,  with two momentum examples $r_\phi=0.1,1$. Bottom:  the $\delta f_\phi$  ratios from different $m_\phi$, with $R_{i/j}$ defined  in Eq.~\eqref{eq:R-ratio}.}
\end{figure*}

We can see that the first term on the right-hand side of Eq.~\eqref{eq:dmudz-fin} corresponds to  the mass effect that drives the scalar into the out-of-equilibrium regime, since in the limit of $m_\phi=0$ the thermal distribution function becomes independent of $z$ and the first term vanishes. The mass  effect becomes more prominent  for small Yukawa couplings and a large $m_\phi$. Indeed,  neither the Yukawa couplings can be too large, nor can the scalar mass be too small. Otherwise the mass effect in the first term would never be significant to create a large $\tilde{\mu}$ from being  zero initially.   Nevertheless,  the Yukawa couplings cannot be too small since the lepton  asymmetry depends on the quartic Yukawa couplings.  Besides, the mass cannot be too large if we simply look at  the overall  mass scaling of the lepton asymmetry  given in Eq.~\eqref{eq:Yell}.

Then the central question arises: what happens to the final lepton asymmetry  if we vary the Yukawa couplings and the mass? While we may write down   the solution of $\tilde\mu(z)$ in the  general analytic  form
\begin{align}\label{eq:general-sol}
	\tilde{\mu}(z)=e^{\int_0^z\mathcal{D} (z')dz'}\int_0^z \frac{z'}{\xi(z')}e^{-\int_0^{z'}\mathcal{D} (z'')dz''}dz'\,,
\end{align}
the dependence on the mass and couplings  will be nontrivially  encoded in the damping force appearing in the  exponent. Therefore,  there is no simple scaling with $|y|^2$ and $m_\phi$  even if a general analytic solution is available, and answering the above  question cannot be achieved by simply looking at the structure of  the Boltzmann equation.

As will be shown in section~\ref{sec:mass-coupling}, the key observation is that the damping force can lead to  an approximate  scaling 
 \begin{align}\label{eq:simple-scaling}
\delta f(z)\propto  \left(\frac{1}{|y|^2}\right) \left(\frac{m}{\bar M_{\rm Pl}}\right)\,,
 \end{align}
 at $z=\mathcal{O}(1)$, which is the very moment when leptogenesis culminates. Consequently, we can infer from  Eq.~\eqref{eq:Yell}   that the variation of $y$ and $m$ will not cause significant changes on $Y_\ell$ when the leptogenesis process is in the most active epoch.

\section{Varying mass and couplings}\label{sec:mass-coupling}

To see the behavior of $\delta f_\phi$ under variation of Yukawa couplings and the scalar mass, we numerically solve the stiff Eq.~\eqref{eq:dmudz-fin} in the two momentum modes $r_\phi=0.1, 1$. 

\begin{figure*}[t]
	\centering
	\includegraphics[scale=0.301]{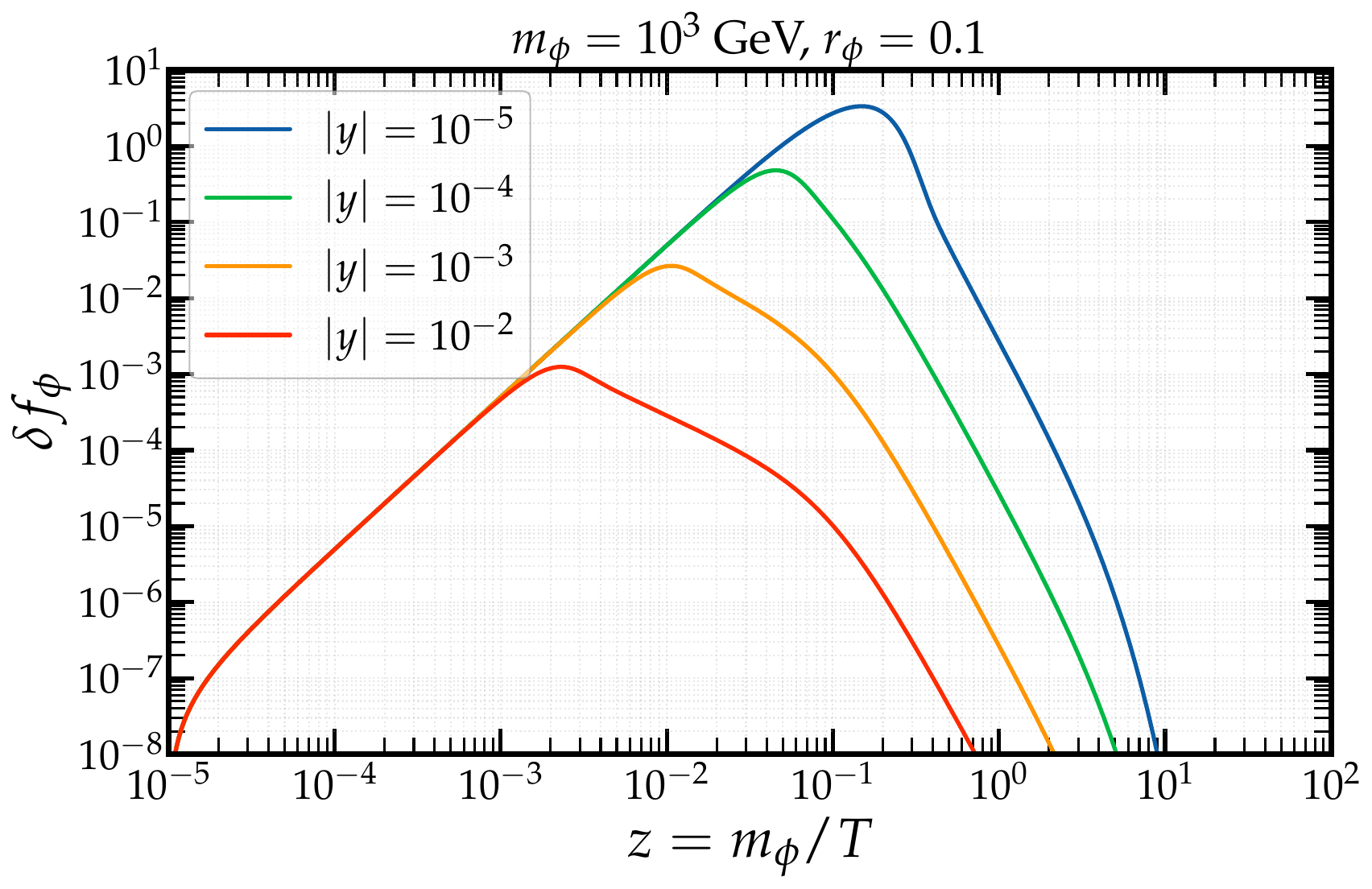} \quad
	\includegraphics[scale=0.301]{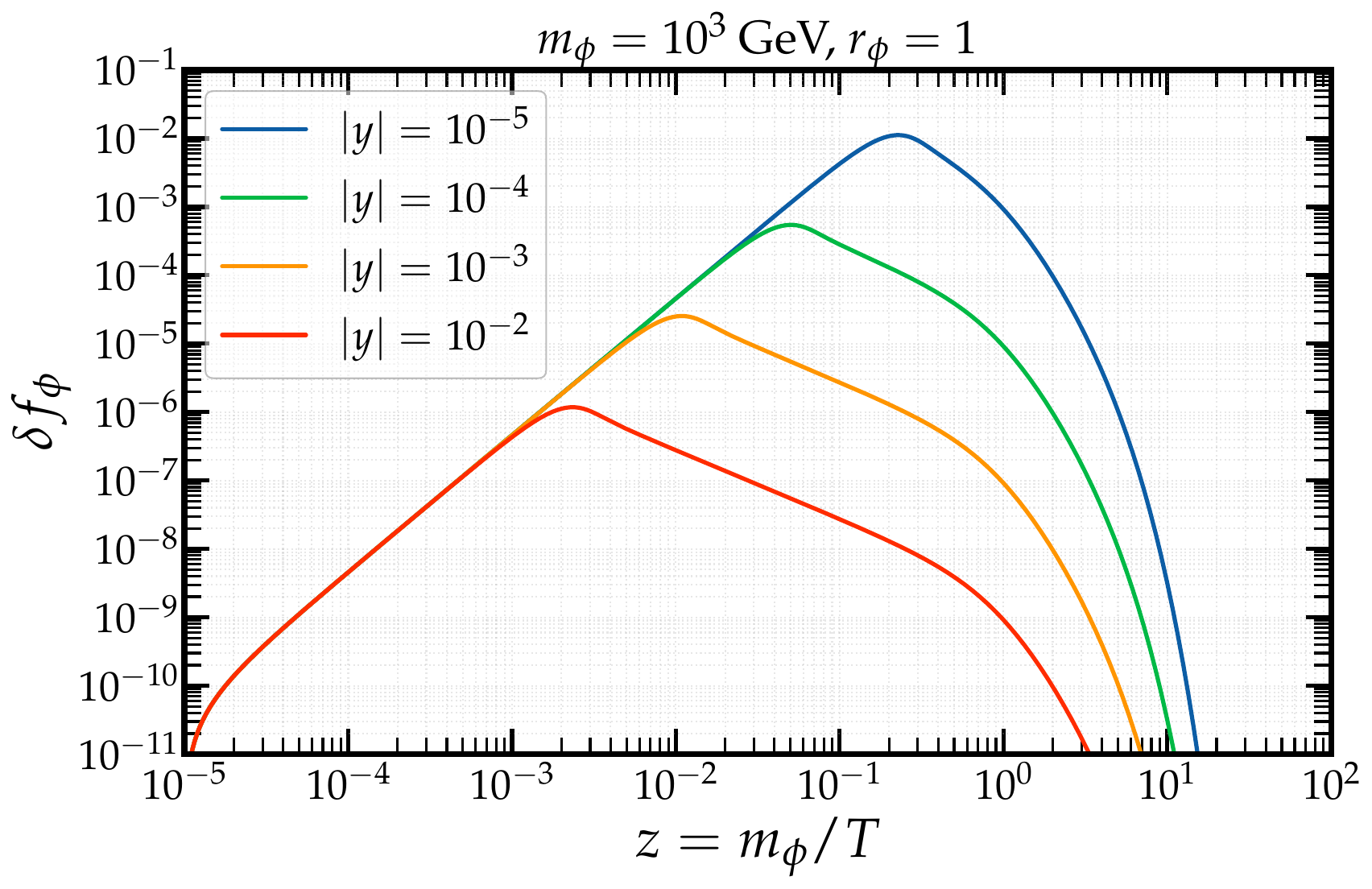}
		\\[0.5cm]
	\includegraphics[scale=0.301]{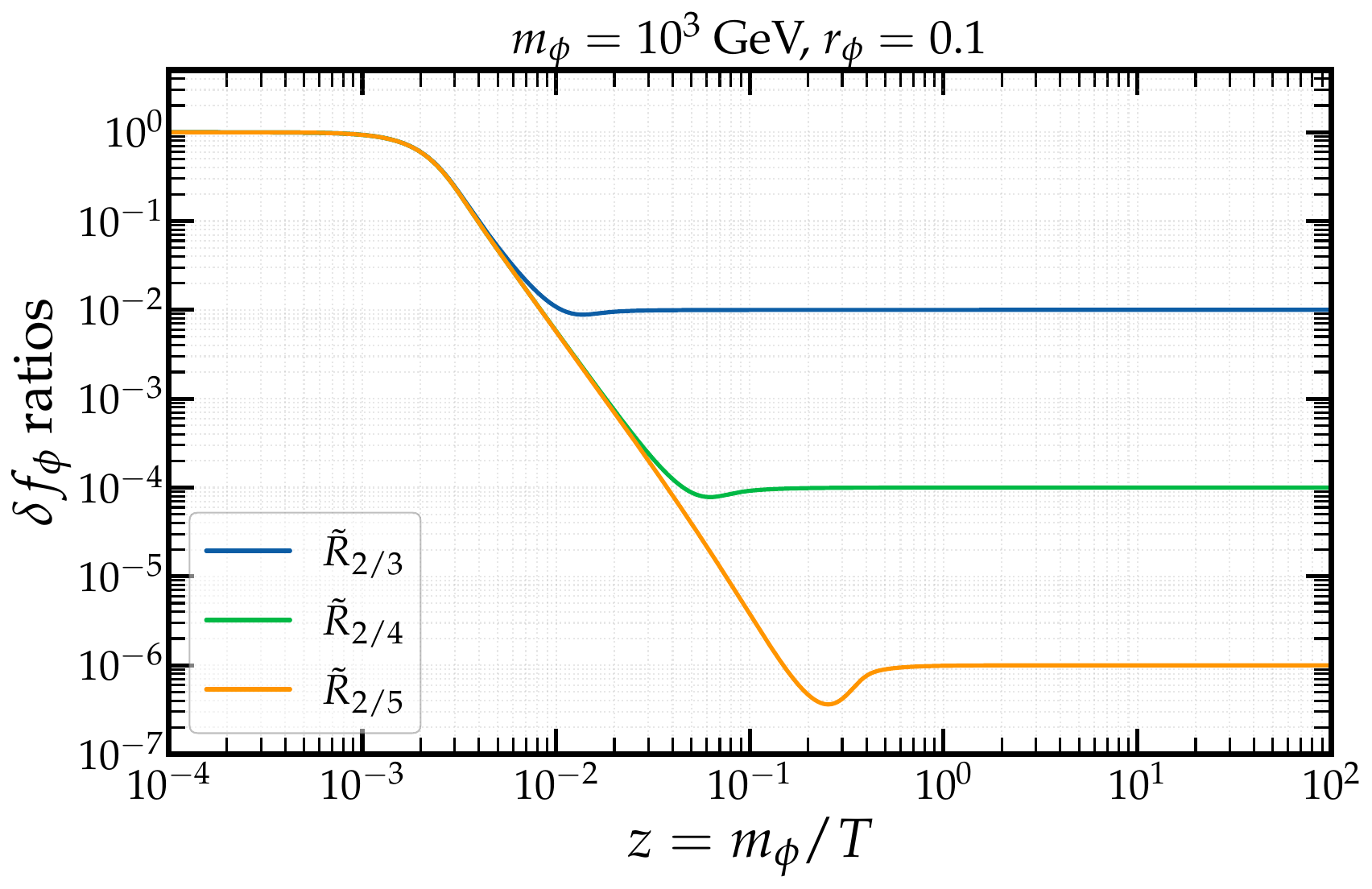} \quad
	\includegraphics[scale=0.301]{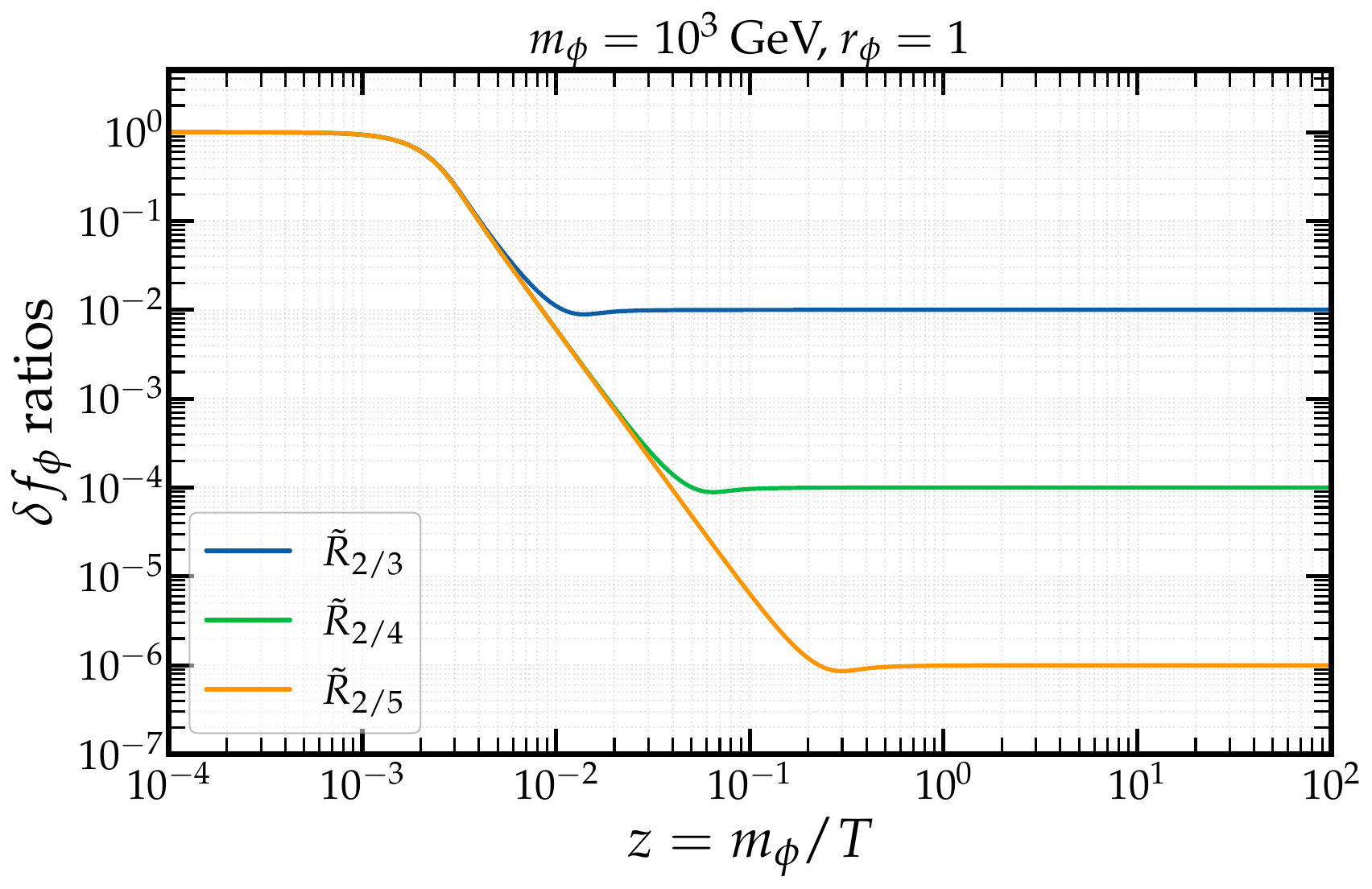}
	\caption{\label{fig:fphi-y}Top: the evolution of $\delta f_\phi$ under the variation of  Yukawa couplings.  The scalar mass  is fixed at $10^{3}$~GeV for reference,  with two momentum examples $r_\phi=0.1,1$. Bottom:  the $\delta f_\phi$  ratios from different $|y|$, with $\tilde R_{i/j}$ defined in Eq.~\eqref{eq:tildeR-ratio}.  }
\end{figure*}

In Fig.~\ref{fig:fphi-m}, the evolution of  $\delta f_\phi(z)$  is shown by varying the scalar mass  and fixing  $|y|=10^{-5}$ (see Eq.~\eqref{eq:ysq}).  It is seen that at initial states where $z\ll 1$, no difference of  $\delta f_\phi$ arises under the variation of $m_\phi$. This is because the evolution of $\delta f_\phi$ is initially determined by  the first term on the right-hand side of  Eq.~\eqref{eq:dmudz-fin}, which drags $f_\phi$ from the equilibrium value. As $\tilde{\mu}$ increases, the damping force comes into play, such that $\tilde{\mu}$ will reach a maximum and then decreases afterwards. In the later decreasing stage, it is the dynamic  competition between the mass dragging effect and the damping force that renders the  evolution of $\tilde{\mu}$  not purely exponential. 

Noticeably, the dependence of $\delta f_\phi$ on $m_\phi$ becomes a simple scaling $\delta f_\phi\propto m_\phi$ at $z\gg 1$, as can be seen in the bottom panel of Fig.~\ref{fig:fphi-m}, where we have defined the $\delta f_\phi$ ratios
\begin{align}\label{eq:R-ratio}
	R_{i/j}\equiv \frac{\delta f_\phi(m_\phi=10^i~\text{GeV})}{\delta f_\phi(m_\phi=10^j~\text{GeV})}\,.
\end{align}
For  infrared-dominated leptogenesis where the generation of  the  lepton asymmetry reaches the maximum in the large $z$ regime, we conclude that  increasing the decaying particle mass can enhance the departure from thermal equilibrium, but such enhancement is canceled out by the overall mass scaling of the lepton asymmetry given in Eq.~\eqref{eq:Yell}. Therefore, varying  the decaying particle mass does not necessarily boost leptogenesis. 

In  Fig.~\ref{fig:fphi-y}, we further show  the evolution of  $\delta f_\phi(z)$  by varying  the Yukawa couplings  and fixing the scalar mass at $10^3$~GeV for reference.  Similarly to what we observed in Fig.~\ref{fig:fphi-m}, there is no difference of $\delta f_\phi$ in the small $z$ regime, $z\ll 1$.  As $z$ increases, $\delta f_\phi$ is lifted to a maximum and then drops. The moment for $\delta f_\phi$ reaching the maximum depends on both $m_\phi$ and $y$. Nevertheless, in the large $z$ limit, $z\gg 1$, we see  from the bottom panel of Fig.~\ref{fig:fphi-y} that the dependence of $\delta f_\phi$ on $y$ reduces to a simple scaling $\delta f_\phi\propto |y|^{-2}$, where we have defined the ratios
\begin{align}\label{eq:tildeR-ratio}
	\tilde R_{i/j}\equiv \frac{\delta f_\phi(|y|=10^{-i})}{\delta f_\phi(|y|=10^{-j})}\,.
\end{align}
 For example,  an increase  of $|y|$ by a factor of 10 will lead to a decrease of  $\delta f_\phi$ by a factor of 100. Such a decrease from the evolution of $\delta f_\phi$ will be compensated for by the overall scaling of the lepton asymmetry, $Y_\ell \propto \text{Im}(y^{\prime 2}y^2)$, as   shown in Eq.~\eqref{eq:Yell}.  Therefore, we conclude that varying  the Yukawa couplings does not necessarily boost leptogenesis. 

It should be mentioned, however,  while the dependence on $y$ is not exactly the same between $\text{Im}(y^{\prime 2}y^2)$ and $\delta f_\phi$ as the former exhibits certain flavor dependence, the order-of-magnitude compensation should be generally  expected unless there is  significant cancellation  among the Yukawa matrix elements, owing to either fine-tuning or accidental  flavor symmetries.

In both Fig.~\ref{fig:fphi-m} and Fig.~\ref{fig:fphi-y}, we have taken two exemplary values  for the scalar momentum $r_\phi=p_\phi/T=0.1, 1$. This is motivated by the fact that when the cosmic temperature drops below the decaying particle mass,  the thermally averaged momentum of the decaying particle is at  $r_\phi= \mathcal{O}(1)$, but   choosing other values of $r_\phi$ does not modify the general picture of the $\delta f_\phi$ difference under the variation of   the  mass and Yukawa couplings. Note that for larger masses and smaller Yukawa couplings,  the simple scaling observed in Fig.~\ref{fig:fphi-m} and Fig.~\ref{fig:fphi-y} may  break down already near $z=1$ for $r_\phi \ll 1$.  Nevertheless, we expect the contribution to the  lepton asymmetry at $r_\phi \ll 1$ is small since the  dominant phase-space  integration of distribution functions mostly   appears  at $E (p)\simeq T$.

\begin{figure*}[t]
	\centering
	\includegraphics[scale=0.3]{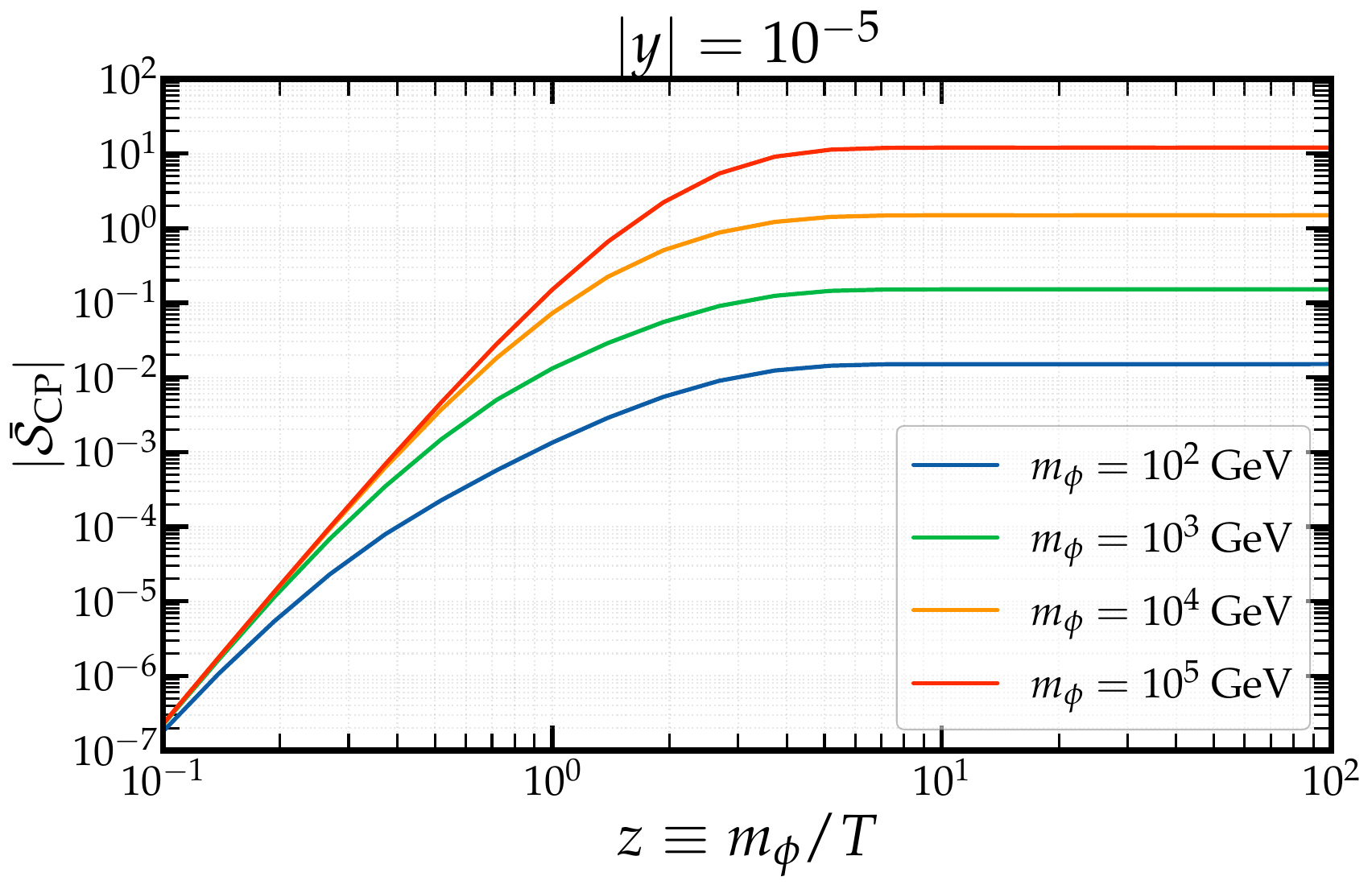} \quad
	\includegraphics[scale=0.3]{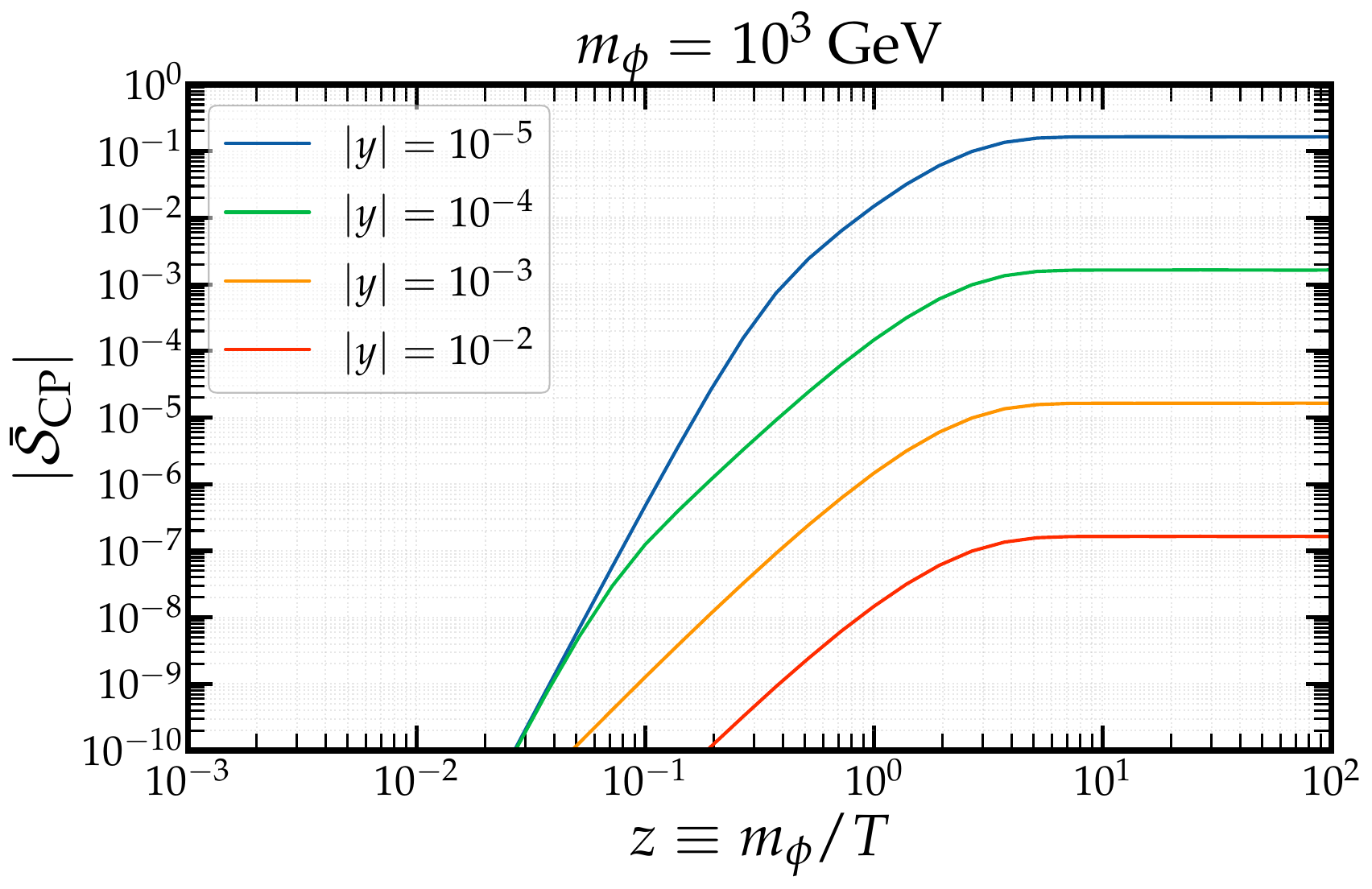}
	\caption{\label{fig:SCP-my} An example of the phase-space  integration over the  CP-violating source $\mathcal{\bar S}_{\rm CP}$ under the variation of the decaying particle mass and Yukawa couplings, respectively.   }
\end{figure*}

\section{Applications}
\label{sec:dis}
To confirm the above  features, now we take an explicit CP-violating source $\mathcal{\bar S}_{\rm CP}$ from Ref.~\cite{Kanemura:2024fbw}
\begin{align}
\mathcal{\bar S}_{\rm CP}=\int_0^\infty \frac{dr_1}{r_1} \int_{r_{\rm low}}^\infty \frac{r_2}{\xi_2} dr_2 \int_{r_{\rm low}}^\infty \frac{r_3}{\xi_3} dr_3 \mathcal{F}\,,
\end{align}
where $r_{\rm low}=z^2/(4x_1)+x_1$ corresponds to the kinematic threshold in particle decay, and   $\xi_i\equiv\sqrt{z^2+r_i^2}$.  $\mathcal{F}$ denotes the product of statistics functions,
\begin{align}
	\mathcal{F}=\delta f_\phi \, f_\phi^{\rm eq}(\xi_2) \left[f_\chi^{\rm eq}(\xi_3-r_1)+f^{\rm eq}_\ell(r_1)-1\right].
\end{align}
We should emphasize  that while  the explicit form of $\mathcal{F}$ is model dependent,  the dominant dependence on the decaying particle mass and Yukawa couplings will  come from the nonthermal distribution function, provided that the overall scaling given in Eq.~\eqref{eq:SCP-deltaf} is a result of  the leptogenesis scenarios. 

We shown in Fig.~\ref{fig:SCP-my} the evolution of $\mathcal{\bar S}_{\rm CP}$ in terms of the dimensionless time variable $z$. We can verify that in the end of leptogenesis, where $z\gtrsim 1$, the variation of  $\mathcal{\bar S}_{\rm CP}$ under different $m_\phi$ and $|y|$ scales as what we expect according to Eq.~\eqref{eq:simple-scaling}, even though such a simple scaling does not appear in the small $z$ regime.

The above results demonstrate that for a generic class of leptogenesis scenarios  where a heavy decaying particle and some decay products contribute to the nonthermal condition in the weak washout regime, variation of the decaying particle mass and Yukawa couplings associated with the thermal decay products does not  boost leptogenesis as naively expected.   The key observation  lies in the simple dependence  of   $\delta f_{\phi}$ on the decaying particle mass  and the Yukawa couplings in the large $z$ regime, which is also the very moment when     leptogenesis  is  most active.

However, we should emphasize  that the cancellation effect is not exact. We can infer from Fig.~\ref{fig:fphi-m} and Fig.~\ref{fig:fphi-y} that  one way to circumvent such cancellation is   early abrupt termination  of leptogenesis at $z<1$ when  $\delta f_\phi$ is still dominated by the $z/\xi$ term in Eq.~\eqref{eq:dmudz-fin}. In this case, varying the mass and Yukawa couplings does not change $\delta f_\phi$ but  it does change  the lepton asymmetry via the imaginary part of Yukawa couplings and the overall  mass scaling shown  in Eq.~\eqref{eq:Yell}.  A second way to avoid the large cancellation  is  flavor effects, as mentioned below Eq.~\eqref{eq:tildeR-ratio}. The dependence of the $\delta f_\phi$ evolution on the Yukawa couplings results from a summation of quadratic Yukawa elements, while the dependence of $\text{Im}[y^{\prime 2}y^2]$ on $y^2$ is encoded in  a  summation of  quartic Yukawa elements. It can render these two kinds of dependence  disentangled if there are significant flavor effects in Yukawa matrix elements, as considered in some low-scale leptogenesis~\cite{Shaposhnikov:2006nn,Gavela:2009cd,Abada:2018oly}.  

Finally, if the weak washout approximation no longer holds, the lepton asymmetry cannot be described by the simple integral equation shown in Eq.~\eqref{eq:Yell}, but  should  be given by the full integro-differential Boltzmann equation. In particular, in the strong washout regime where the late-time evolution of the lepton asymmetry is important,  the dependence of the final lepton asymmetry on the mass and Yukawa couplings will become more complicated, yet  the  generalization to the strong washout regime deserves consideration.

\section{Conclusion}
\label{sec:con}
We considered  a class of  weak-washout leptogenesis scenarios in this work, where heavy particle decay generates the CP asymmetry and the nonthermal conditions are provided by the decaying particle and some decay products. We have demonstrated a general feature in terms of the mass and coupling effects. Based on the general parameterization given in Eq.~\eqref{eq:Yell}, we found that varying the Yukawa couplings and the decaying particle mass does not necessarily boost leptogenesis, contrary to  naive expectations. 

Without relying on numerical parameter scans of a given particle physics model,  the common feature implies that any realization of such a class  of leptogenesis from a particular benchmark point will automatically open a much broader parameter space.  In turn, if   leptogenesis cannot be realized by a given set of parameters,  boosting leptogenesis would then be challenging by simply tuning the orders of magnitude for  the  Yukawa couplings and the decaying particle mass,  unless leptogenesis terminates at earlier epochs when the cosmic temperature is still higher than the decaying particle mass, or there exist significant flavor effects in Yukawa matrix elements.

\section*{Acknowledgements}
This project is supported by JSPS Grant-in-Aid for JSPS Research Fellows No. 24KF0060. SK is also supported in part by Grants-in-Aid for Scientific Research(KAKENHI) Nos. 23K17691 and 20H00160.

\bibliographystyle{JHEP}
\bibliography{Refs}

\end{document}